# Three Futures for the Diagnostic Radiologist: A Structured Disagreement About What AI Actually Changes

*Viewpoint*

Jan Beger[1] · Amine Korchi, MD[2] · Christoph A. Agten, MD[3]

[1] Global Head of AI Advocacy, GE HealthCare; Executive Director, HelloAI · [2] Radiologist and Neuroradiologist · [3] Consultant Radiologist, Musculoskeletal Radiology

*Correspondence: Jan Beger · jan.beger@me.com · janbeger.ai*

---

**ABSTRACT**

**Rationale.** The diagnostic radiologist's role in 2035 will not look like it does today. Imaging AI is already changing how worklists are organized, how reports are generated, and which cases require a radiologist's attention. What remains genuinely contested is not whether the role changes but how.

**Approach.** Three subject-matter experts (two radiologists and one health tech professional with more than 20 years of experience in medical imaging IT) independently authored 2035 job descriptions for the diagnostic radiologist using a shared template. Each author wrote from a distinct vantage point: one optimistic, one framed as a trade-off view incorporating workforce economics, and one structured around professional stratification. The three versions were published openly and subjected to a structured comparison across seven dimensions.

**Key findings.** The three versions agree on direction but disagree on magnitude. All three describe a radiologist whose routine workload is AI-managed, who carries accountability for AI output, and who spends more time on complex cases and clinical collaboration than today's radiologist does. They diverge on headcount, career security, and whether the profession expands broadly, concentrates into a smaller well-compensated group, or stratifies into sharply differentiated tiers.

**Conclusion.** AI won't eliminate the diagnostic radiologist. Whether it expands, concentrates, or stratifies the profession depends on choices health systems haven't made yet. The clinical argument for optimism is real. So is the economic argument for caution. Both can be true simultaneously.

---

## 1. Introduction and rationale

The question of what diagnostic radiologists will do in 2035 gets asked constantly and answered poorly. Two scenarios dominate the conversation. The first is technological optimism: AI expands the profession, radiologists become orchestrators, everyone wins. The second is simpler: AI replaces routine reads, headcount falls, the job shrinks. Both framings contain real evidence. Neither is complete on its own.

This paper documents a deliberate experiment in structured disagreement. Three people with different vantage points on the same problem independently wrote the 2035 job description for a diagnostic radiologist. Two are practising radiologists with active clinical AI experience. One is a health tech professional with more than 20 years of experience in medical imaging IT, working across AI deployment, strategy, and health system integration. Each author used a shared template (position summary, key responsibilities, required qualifications, work environment, and success measures) but wrote the content independently, without coordination. The three job descriptions were published as a public resource (janbeger.ai/radiologist-2035/), structured to allow direct comparison across seven dimensions, and subjected to an open vote from the radiology community.

The goal was not a consensus. It was clarity about where the disagreements actually sit and what evidence bears on each of them. This paper presents the three scenarios, characterizes where they diverge, and draws out the implications for training, governance, and workforce planning.

## 2. The three 2035 scenarios

### 2.1. The optimistic scenario (Korchi)

Dr. Korchi's vision describes a radiologist empowered by AI, leading its development and adoption, emerging as a leading specialty in diagnostics, and ultimately becoming a diagnostic orchestrator. AI manages speed and initial detection, while expanding radiologists' diagnostic capacity through greater depth (novel biomarkers, quantitative insights, and prognostic information) and greater breadth (higher volumes, broader scope of practice, and more integrated diagnostic oversight). The radiologist retains full authority over outputs, leads multidisciplinary discussions, and expands into direct patient communication that today's reading-room model rarely supports. The profession grows: AI generates new imaging value, and radiologists grow with it. The orchestrator framing is not new. Jha and Topol described it as "information specialist" in 2016 [1], and the language has carried forward in the perspective literature since, most recently as "diagnostic orchestrator" and "superdiagnostician." Korchi's version is the strongest current articulation of that thread.

The job in this version is better than today's in almost every respect. More complex cases. More added value. More clinical visibility. More direct relationships with referring physicians and patients. More leadership across the clinical team. The technical requirement for AI fluency is added, but it is framed as an opportunity for expansion rather than a defensive necessity.

The core claim is that AI's effect on imaging demand outweighs its effect on per-radiologist throughput.

### 2.2. The trade-off scenario (Beger)

Beger's version accepts the clinical argument in Korchi's description but adds an economic constraint that the optimistic version underweights. More output per radiologist means fewer radiologists per department. That is not a failure of the profession. It is how automation works in every industry where it has been deployed at scale, and there is no structural reason to expect radiology to be an exception.

In this version, the job that reaches the 2035 radiologist is genuinely harder. Routine studies are AI-managed before they arrive at the reading desk. What lands there is ambiguous, rare, or high-stakes. The radiologist carries accountability for direct interpretation and for AI-processed reports validated under their name. That is a real clinical function and a heavier one than it sounds.

The consultative role expands for the same reason. Less time on routine volume, more time in multidisciplinary meetings, direct patient communication, and clinical conversations with referring physicians. The 2035 radiologist in this version is more visible and more exposed. The economics of the shift means fewer radiologists hold this better job. The version explicitly notes that regulatory timelines, reimbursement frameworks, and institutional restructuring slow the pace of change, making 2035 a direction rather than a deadline.

### 2.3. The stratification scenario (Agten)

Dr. Agten's version, labeled the pessimistic view on the public site, does not describe one job. It describes a profession that splits into three tiers. The first is a scarce, highly compensated subspecialist layer that handles complex edge cases and holds governance authority over the AI pipeline. The second is a hands-on procedural layer, valuable precisely because AI cannot physically perform ultrasound or guide a needle. The third is a large, remote audit workforce reviewing AI output at high volume for compensation that sits just above resident-level pay.

The structural logic of this version is supply and demand. If the number of radiologists who can do routine reads is large and the demand for them falls as AI automates those reads, compensation and conditions in that segment deteriorate. The elite tier remains scarce and well-compensated because complexity is hard to automate. The procedural tier is protected by physical presence. The audit tier forms because autonomous AI still needs human sign-off, and that human does not need to be at the top of the pay scale.

This version is the most uncomfortable to read and the most important to take seriously. The clinical argument is real; so is the economic stratification argument. Both can be true simultaneously.

## 3. Comparative discussion

### 3.1. Key areas of agreement and disagreement

Table 1 shows the structured comparison across seven dimensions. The versions agree on two things: AI autonomy on routine reads increases substantially by 2035, and the radiologist's role as the accountable clinical authority persists. Everything else is contested.

| Dimension | Korchi (Optimistic) | Beger (Trade-off) | Agten (Pessimistic / Stratification) |
|---|---|---|---|
| **AI autonomy on routine reads** | High (co-pilot model) | High (AI drafts; radiologist validates) | Very high (autonomous; humans audit) |
| **Radiologist clinical authority** | Full, active | Full, final accountability | Tiered: full only at Tier 1 |
| **Direct patient contact** | Increased, deliberate | Increased, by necessity | Rare except Tier 1 and Tier 2 |
| **Interventional & procedural work** | Expanded | Retained; procedural value grows | Core for Tier 2; absent for Tier 3 |
| **AI governance load** | High (AI stewardship) | High (post-market surveillance) | High for Tier 1; minimal for Tier 3 |
| **Radiologist headcount** | Grows with imaging demand | Falls: more output per radiologist | Falls and stratifies |
| **Career security** | Strong across the profession | Strong for those who hold the job | High for elite; low for Tier 3 |
| **Overall verdict** | **Expansion** | **Concentration** | **Stratification** |

**Table 1.** Structured comparison of the three 2035 radiologist scenarios across seven dimensions. All columns are derived from the authors' full texts. Agten's column reflects conditions across all three tiers; individual tiers differ substantially.

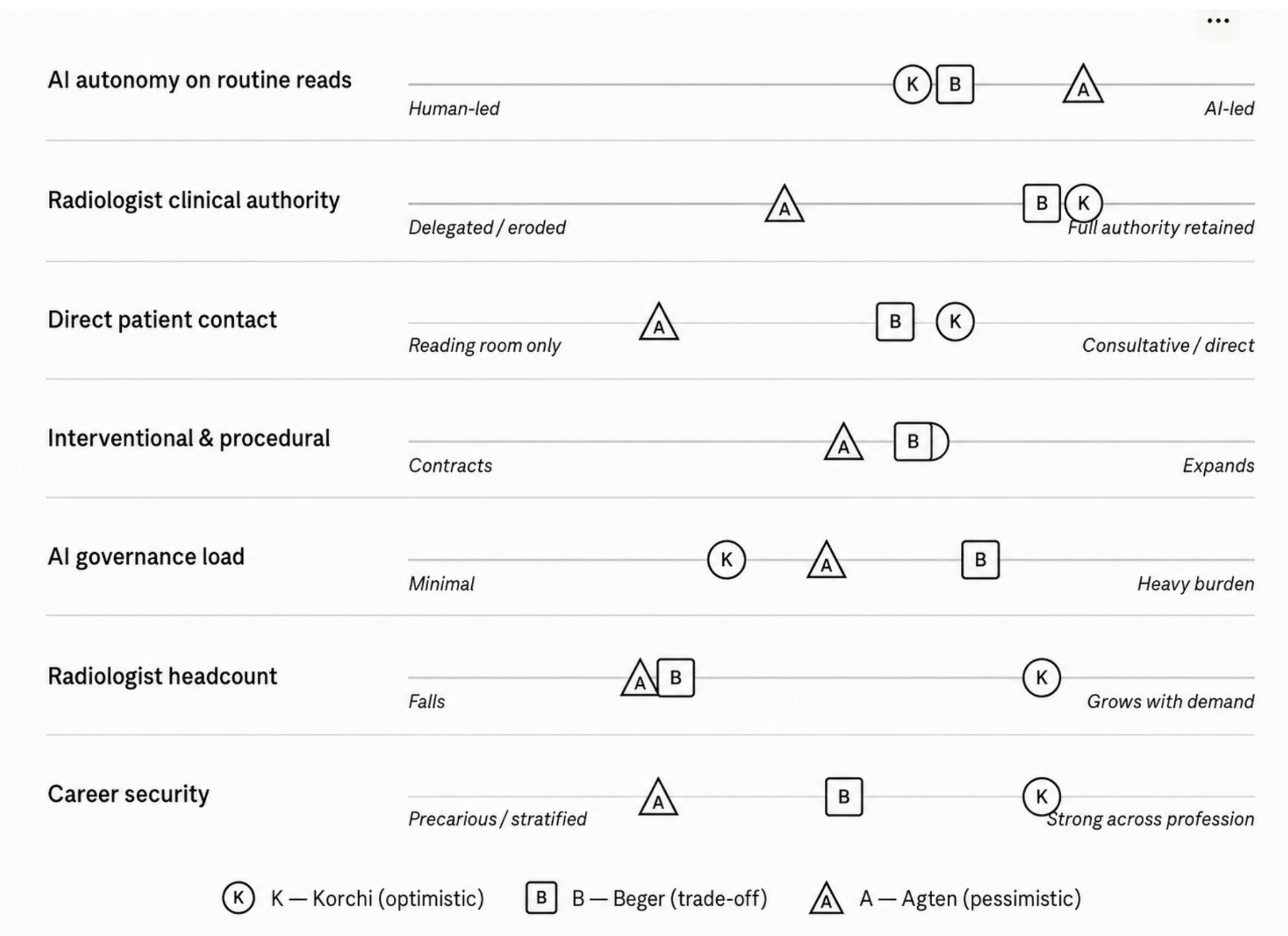

**Figure 1.** Author positions across seven dimensions of the 2035 radiologist role. Each marker shows the author's position on a spectrum from human-dominant to AI-dominant outcome. Marker spread on a given dimension indicates genuine disagreement; marker clustering indicates convergence. Positions are derived from the authors' full published texts (janbeger.ai/radiologist-2035/).

The headcount dimension is where the practical stakes are highest. Korchi's version predicts a larger profession. The trade-off scenario predicts a smaller one, with better conditions for those who hold the jobs. Agten's version predicts stratification: the total radiologist pool may not shrink dramatically, but the distribution of pay, conditions, and career quality becomes highly unequal.

The governance dimension deserves specific attention. All three versions assign meaningful governance responsibility to radiologists. In Korchi's version, this is framed as AI stewardship: radiologists as the accountable owners of the AI pipeline across the full diagnostic services, not just radiology. In the trade-off scenario, it is post-market surveillance and drift detection, with legal accountability for AI-processed reports signed under the radiologist's name. In Agten's version, governance sits at Tier 1 and does not exist at Tier 3. The regulatory implications of this divergence are not trivial: if accountability for AI output is assumed to rest with a licensed radiologist, the question of which radiologist and under what conditions matters for liability frameworks.

### 3.2. What the evidence supports

Several claims across the three versions are grounded in converging evidence. The MASAI trial found that AI-supported single reading produced equivalent diagnostic performance to standard double reading with a 44% reduction in screen-reading workload [2,3]. An independent UK trial replacing the second human reader with AI in NHS breast screening found the same: equivalent performance, 46% less human reading [4]. Both trials support the trade-off scenario. Neither supports full autonomous AI reporting; a human radiologist remained accountable throughout in both. Neither resolves the headcount question, which depends on whether efficiency gains are absorbed as cost savings or redirected into expanded capacity.

AI triage tools show meaningful reductions in time-to-diagnosis [5,6], but whether that translates into redistribution of radiologist attention toward complex work remains undocumented. The economic argument that AI substitutes routine labor rather than augments it has been made explicitly in the peer-reviewed literature [7], and the structural pressure is real: the UK faces a 30% shortfall of radiologists against rising demand [8]. Reimbursement incentives have already shaped deployment: financial logic, not diagnostic yield, determines which AI enters practice first [9].

The stratification scenario has the least direct empirical support today, since large-scale autonomous AI reporting isn't deployed at the scope Agten describes. Its logic is consistent with the documented pattern of automation concentrating wage gains among workers whose tasks can't be automated [10].

### 3.3. What none of the scenarios can resolve

All three versions are built on assumptions about regulatory timelines, reimbursement structure, and institutional restructuring that are difficult to predict. Healthcare AI deployment moves slower than the underlying technology. Regulatory approval, post-market surveillance requirements, and liability frameworks for AI-generated reports are still taking shape across jurisdictions [11]. The European AI Act's classification of high-risk AI systems in medical imaging, for example, imposes requirements on manufacturers and users that will shape how quickly autonomous AI reporting can be deployed at scale [12]. The directional pattern of reimbursement distortion is already documented: financial incentives have skewed which AI tools enter practice first, but the magnitude and jurisdiction-specific consequences remain uncertain and will vary substantially across health systems [9]. Even where approved AI tools are running in clinical environments, continuous performance monitoring remains a documented weakness: most radiology departments currently lack the staff and infrastructure to sustain meaningful post-deployment evaluation of the AI systems they operate [9].

The deskilling question also cuts across all three versions. If radiology trainees spend fewer hours reading routine studies because AI handles those reads, the clinical judgment built through volume may not develop in the same way. The optimistic version assumes AI fluency substitutes for this and even adds value. The trade-off scenario notes the concern without resolving it. Agten's version builds it into the Tier 3 structure: auditors reviewing AI output are doing a qualitatively different activity than reading de novo. Research on clinical trainees using AI identifies three distinct failure modes: never-skilling (skipping fundamentals before they form), deskilling (eroding capability that already exists), and mis-skilling (building on flawed AI outputs) [13]. Whether any of these failure modes apply at specialty training scale is an open question for radiology training programs and certification bodies.

A separate uncertainty sits one layer deeper. Multimodal AI does not only change what the radiologist does; it changes the information environment the radiologist works inside. Fusing imaging with genomics, pathology, EHR data, and wearable signals into a single synthesised decision view shifts the cognitive task from direct perceptual engagement with images to interpreting pre-digested outputs the radiologist did not generate, and may not have the training to fully interrogate. Radiologists have deep expertise in medical imaging interpretation. They have no formal training in genomics (a domain where even the clinical vocabulary is unfamiliar), and their exposure to histopathology is limited to what cross-disciplinary collaboration provides. All three of our scenarios assume the radiologist absorbs whatever multimodal output the system produces. None of them treat that absorption as a separate variable. It is one.

A separate but related concern applies to practicing radiologists. AI assistance does not augment all radiologists equally. A large experimental study of 140 radiologists across 15 chest radiograph pathologies found substantial heterogeneity in treatment effects: some radiologists performed meaningfully worse with AI than without it, and conventional proxies such as years of experience, subspecialty, and prior AI familiarity did not predict who would benefit [14]. This challenges the assumption, implicit in all three versions, that AI assistance is a uniform performance lift. The actual benefit depends heavily on AI accuracy in the specific deployment context and on individual variation that current tools cannot anticipate.

### 3.4. What the public response revealed

The three job descriptions were published at janbeger.ai/radiologist-2035/ with a five-question self-routing test. Visitors answer questions about their assumptions on AI autonomy, headcount, and career security, and the test routes them to the version that most closely matches their outlook. No demographic data are collected, responses are self-selected, and the test is designed for engagement rather than statistical inference. The observations below draw on LinkedIn comment threads across four posts: over 100 reactions and 60 comments from radiology and health tech communities in the first week.

The distribution was not uniform. Of the named test-takers who shared results publicly, the trade-off scenario attracted the most responses, with optimistic and pessimistic roughly splitting the remainder. Readers who reported active AI deployment in daily practice (a health AI regulation consultant, a commercial radiology platform operator, a breast interventional radiologist) skewed trade-off or pessimistic. Those without hands-on implementation experience skewed optimistic.

Two objections surfaced repeatedly. Regulatory frameworks were cited as the strongest structural constraint on the pessimistic scenario: mandatory radiologist sign-off under current EU AI Act provisions means the audit tier Agten describes needs a regulatory shift before it can operate at scale. And the Jevons paradox came up as a direct challenge to the headcount-reduction logic: PACS and voice recognition became volume rather than recovered time, and AI may follow the same curve. Neither objection resolves the core disagreement, but both are live uncertainties the three versions don't fully address.

The version a reader finds most credible correlates with their assumptions about economics, not just clinical confidence in AI. The disagreement is partly empirical (what will AI actually be able to do) and partly structural: how will healthcare economics respond to what AI can do. The clinical questions are closer to resolution. The economic ones aren't.

## 4. Synthesis and implications

### 4.1. For training programs

All three versions agree that AI fluency is now a required competency, not an optional one. The EU AI Act Omnibus proposed softening the AI literacy obligation from a hard requirement to a support measure, but deployers of high-risk AI systems, which includes medical imaging AI, retain human oversight training obligations. The practical argument stands independent of the regulatory one: a radiologist who can't interrogate AI output is professionally exposed whether or not the law demands otherwise.

Where they disagree is what that means in practice. Korchi's version emphasizes co-development and clinical leadership. The trade-off scenario emphasizes post-market surveillance, drift detection, and the specific competencies required to carry legal accountability for AI-validated reports. Agten's version implies that Tier 1 radiologists need governance and adjudication skills, while Tier 3 auditors need something closer to quality-control methodology.

Training programs that prepare radiologists for a single version of the future are taking a position on which scenario will obtain. A more defensible approach is to treat the versions as a skills map: the competencies needed in the optimistic scenario (orchestration, clinical leadership, patient communication) are valuable even in the stratification scenario, particularly for those who want to land in Tier 1 rather than Tier 3. Subspecialty depth and procedural competence are explicitly protective in all three scenarios.

### 4.2. For workforce planners

Workforce planning in radiology has historically assumed a linear relationship between imaging volume and radiologist headcount. AI breaks that assumption. The trade-off and stratification scenarios both predict that efficiency gains decouple headcount from volume. Health systems that plan on the optimistic scenario and find themselves in the trade-off scenario will face structural mismatches between supply and demand.

The practical implication is not to assume the most conservative scenario, but to plan for a range. The structural factors that determine which scenario obtains (reimbursement for AI-assisted reads, liability frameworks for AI sign-off, training pipeline for subspecialty depth) are mostly outside individual radiologists' control but within the influence of professional societies, certifying bodies, and health system leadership.

### 4.3. For AI vendors and developers

The three scenarios make different implicit demands on AI developers. The optimistic scenario requires AI that genuinely functions as a co-pilot: controllable, transparent, and deferential to radiologist judgment. The trade-off scenario requires AI that is reliable enough to be signed off at scale, with auditable performance data that supports the governance functions radiologists are expected to perform. The stratification scenario requires AI reliable enough to run autonomously on high-confidence routine studies, which is a substantially higher bar than co-pilot deployment.

Current imaging AI is closer to the co-pilot end of that spectrum than the autonomous end. Closing the gap requires validation approaches that reflect real-world deployment conditions, not curated retrospective datasets. It also requires attention to the accountability question: if a radiologist is signing off on AI-processed reports at scale, what tooling does the AI system provide to support that function, and what happens when the AI is wrong in ways the radiologist could not reasonably have caught?

### 4.4. Conclusion

We do not agree with each other, and that is the point.

The three versions presented here are not three opinions about the same set of facts. They are three positions about which facts matter most. The clinical case for the optimistic scenario is real. The economic case for the trade-off scenario is real. The structural logic of the stratification scenario is real. Anyone who tells you the future of this job is settled is suppressing one of these three lines of argument.

What we can say with confidence is narrower but more useful. The routine read is contracting. Complex work is concentrating. Direct clinical relationships are expanding. AI governance is becoming a core radiologist function, not a side responsibility. And the headcount math, which has historically tracked imaging volume, is being decoupled from it by per-radiologist efficiency gains.

For a radiologist entering the field today, planning for only one of these scenarios is a position on which argument will win. The safer approach is to build competencies that are valuable regardless of which scenario obtains: subspecialty depth, procedural skills, AI fluency, and the clinical relationships that AI cannot replicate. The job in 2035 will be different from today's, and probably better for the people who hold it. Whether the radiology workforce of 2035 will be as large as it is today is the question the field has not yet answered honestly. This paper does not answer it either, but it maps where the disagreement actually sits, which is a necessary first step.

## Declarations

Conflict of interest: Jan Beger is employed by GE HealthCare as Global Head of AI Advocacy. GE HealthCare develops and markets imaging AI products. The analysis and positions expressed here are the author's own and do not represent GE HealthCare's views. Amine Korchi is founder and director of Singularity Consulting SARL. He holds shares in Aeon, Butterfly Network, Cerebriu, Domo Health, GE HealthCare, Gleamer, Hyperfine, Microsoft, Nvidia, Oracle, Primaa, RadNet, Sophia Genetics, ThinkSono, and Vocca. He advises or consults for Aeon, Bonescreen, Cerebriu, Computer Vision AG, CSS, Future of Health Grants, Gleamer, HCK, Hyperfine, Mediaire, the Swiss Federal Institute of Technology, and Innovation Park Foundation. He receives speaker fees from GE HealthCare. Christoph Agten declares no conflicts of interest.

Funding: None.

Author contributions: All three authors independently wrote complete 2035 job descriptions, which form the primary source material for this paper. Beger led synthesis, structural comparison, and manuscript preparation. Korchi and Agten reviewed and approved the final text.

Ethics: Not applicable.

Data availability: Full text of all three job descriptions and comparison materials are publicly accessible at https://janbeger.ai/radiologist-2035/